# Evaluation of a general model for multimodal unsaturated soil hydraulic properties[1]

Katsutoshi Seki[1*], Nobuo Toride[2], Martinus Th. van Genuchten[3,4]

[1]Natural Science Laboratory, Toyo University, 5-28-20 Hakusan, Bunkyo-ku, Tokyo 112-8606, Japan.
[2]Graduate school of Bioresources, Mie University, 1577 Kurimamachiya-cho Tsu, Mie 514-850, Japan.
[3]Department of Earth Sciences, Utrecht University, Princetonlaan 8a, 3584 CS, Utrecht, Netherlands.
[4]Department of Nuclear Engineering, POLI & COPPE, Federal University of Rio de Janeiro, UFRJ, Rua Horácio Macedo, Bloco G, Cidade Universitária, Rio de Janeiro, RJ 21941-450, Brazil.
[*]Corresponding author. E-mail: seki_k@toyo.jp

**Abstract:** Many soils and other porous media exhibit dual- or multi-porosity type features. In a previous study (Seki et al., 2022) we presented multimodal water retention and closed-form hydraulic conductivity equations for such media. The objective of this study is to show that the proposed equations are practically useful. Specifically, dual-BC (Brooks and Corey)-CH (common head) (DBC), dual-VG (van Genuchten)-CH (DVC), and KO (Kosugi)$_1$BC$_2$-CH (KBC) models were evaluated for a broad range of soil types. The three models showed good agreement with measured water retention and hydraulic conductivity data over a wide range of pressure heads. Results were obtained by first optimizing water retention parameters and then optimizing the saturated hydraulic conductivity ($K_s$) and two parameters ($p$, $q$) or ($p$, $r$) in the general hydraulic conductivity equation. Although conventionally the tortuosity factor $p$ is optimized and ($q$, $r$) fixed, sensitivity analyses showed that optimization of two parameters ($p+r$, $qr$) is required for the multimodal models. For 20 soils from the UNSODA database, the average $R^2$ for log (hydraulic conductivity) was highest (0.985) for the KBC model with $r$=1 and optimization of ($K_s$, $p$, $q$). This result was almost equivalent (0.973) to the DVC model with $q$=1 and optimization of ($K_s$, $p$, $r$); both were higher than $R^2$ for the widely used Peters model (0.956) when optimizing ($K_s$, $p$, $a$, $\omega$). The proposed equations are useful for practical applications while mathematically being simple and consistent.

## INTRODUCTION

Knowledge of the water retention (WRF) and hydraulic conductivity (HCF) functions is important for simulating water retention and flow in unsaturated soils and other porous media. Frequently used analytical expressions for these functions were proposed by Brooks and Corey (1964), van Genuchten (1980) and Kosugi (1996), further referred to as the BC, VG and KO models respectively. The models were obtained by combining expressions for the water retention function with statistical pore-size distribution models that consider soils to be made up of a bundle of capillary tubes (Burdine, 1953; Mualem, 1976).

Formally, the above models are restricted to capillary flow within a bundle of capillary tubes. As the water content of a soil decreases, the retention of water changes from capillarity to fluid adsorption, with the mechanism of water flow changing from capillary movement to film and corner flow. For this reason, several used Mualem's model only at relatively high water contents (the capillary water range) and modeled the hydraulic conductivity at low water content (i.e., the adsorption water range) independently (e.g., Peters, 2013), or avoided the use of capillary bundle models altogether to independently describe the WRF and HCF over the whole moisture range (e.g., Luo et al., 2022).

Another approach for expressing the HCF over a wide range of pressure heads is to make the Burdine or Mualem expressions more flexible by increasing the number of adjustable parameters. Hoffmann-Riem et al. (1999) proposed for this purpose a general HCF with three adjustable parameters $p$, $q$, $r$ as

$$K_r(h) = \frac{K(h)}{K_s} = S(h)^p \left[ \frac{\int_0^{S(h)} h(S)^{-q} dS}{\int_0^1 h(S)^{-q} dS} \right]^r \qquad (1)$$

where $h$ is the pressure head (assumed in this study to be positive for unsaturated conditions), $K_r$ is the relative hydraulic conductivity, $K$ the unsaturated hydraulic conductivity, $K_s$ the saturated hydraulic conductivity, and $S$ effective saturation defined by

$$S = \frac{\theta - \theta_r}{\theta_s - \theta_r} \qquad (2)$$

where $\theta$ is the volumetric water content, and $\theta_r$ and $\theta_s$ are the residual and saturated water contents, respectively. In particular, $p=2$, $q=2$, and $r=1$ for Burdine's model, and $p=0.5$, $q=1$, and $r=2$ for Mualem's model.

Although the Burdine and Mualem models consider only one parameter (i.e., the tortuosity factor $p$) to be variable, and $q$ and $r$ to be constants, Hoffmann-Riem et al. (1999) showed that optimizing $p$ and $r$ for the VG model would improve estimations of the HCF. They suggested that neither the general formulation nor the simpler Burdine and Mualem variants should be interpreted as being physically based. Kosugi (1999) showed that optimizing two parameters (i.e., $p$ and $q$) improved estimations of the HCF using his KO model, especially when the range of pressure head was wide.

To make the WRF more flexible for multi-porosity media, Durner (1994) proposed a linear superposition of VG equations (the multi-VG retention model). Priesack and Durner (2006) subsequently provided a closed-form hydraulic conductivity equation for the multi-VG model by using the general HCF. The multimodal model proved to be far more flexible in describing the hydraulic properties of both near-saturated soils and dry soils. Mualem's model ($q=1$, $r=2$) with a free parameter $p$ has since been used for multimodal WRFs in many studies (Dimitrov et al., 2014; Lipovetsky et al., 2020; Schelle et al., 2010; Watanabe and Osada, 2016).

In our previous study (Seki et al., 2022) we derived closed-form general HCF equations for the multimodal model in terms of a linear superposition of any combinations of the BC, VG, and KO models. We also showed that the asymptotic slope of log $h$ – log $K$ curve at low water contents (i.e., slope of the subcurve expressing film flow) for Mualem's model cannot be smaller than $qr=2$, which contradicts the result by Peters (2013) that the slope is approximately 1.5 for many soils. We hence suggested that the value of $q$ or $r$ may need to be modified from Mualem's parameter to enhance better descriptions of $K(h)$ at relatively low water contents. Based on this result and recommendations by Hoffmann-Riem et al. (1999) and Kosugi (1999) to consider two parameters of the general HCF as free variables, the hypothesis of this study is that the general HCF with two free HCF parameters $(p, q)$ or $(p, r)$ is able to express the HCF over a wide range of pressure heads.

In this study we apply the general HCF equations of the multimodal models proposed by Seki et al. (2022) to previously published water retention and hydraulic conductivity data of soils with various textures over a wide range of pressure heads. Our focus is especially on the benefit of optimizing $q$ or $r$ in addition to $p$ in Eq. (1).

**METHODOLOGY**
**Model description**

The water retention function (WRF) of multimodal soils may be defined as

$$S(h) = \sum_{i=1}^{k} w_i \, S_i(h) \qquad (3)$$

where $k$ is the number of subsystems, and $w_i$ are weighing factors with $0 < w_i < 1$ and $\Sigma w_i = 1$. The sub-retention functions $S_i(h)$ used in this study are summarized in Table 1. The closed-form hydraulic conductivity equations derived from $S_i(h)$ in Table 1 with the general HCF equation (Eq. 1) are provided by Seki et al. (2022). Note that the parameter $q$ in the HCF is common with the WRF in the VG subfunction (Table 1). For the multimodal models we assume $\theta_r=0$, which allows Eq. (2) to be written as

$$\theta(h) = \theta_s S(h) \qquad (4)$$

**Table 1**. Sub-retention functions for the Brooks-Corey (BC), van Genuchten (VG) and Kosugi (KO) models.

| Type | Equation | Parameters |
|---|---|---|
| BC | $S_i(h) = \begin{cases} \left(\dfrac{h}{h_{b_i}}\right)^{-\lambda_i} & (h > h_{b_i}) \\ 1 & (h \leq h_{b_i}) \end{cases}$ | $h_{b_i}, \lambda_i$ |
| VG | $S_i(h) = [1 + (\alpha_i h)^{n_i}]^{-m_i}$, $m_i = 1-q/n_i$ | $\alpha_i, n_i, q$ |
| KO | $S_i(h) = Q\left[\dfrac{\ln(h/h_{m_i})}{\sigma_i}\right]$, $Q(x) = \dfrac{1}{2}\left[\mathrm{erfc}\left(\dfrac{x}{\sqrt{2}}\right)\right]$ | $h_{m_i}, \sigma_i$ |

In this paper, we focus on multimodal models with $k=2$, where $S_1(h)$ is expected to express capillary water and $S_2(h)$ adsorbed water. Eqs. (3) and (4) can then be summarized as

$$\theta(h) = \theta_s[w S_1(h) + (1-w)\, S_2(h)] \qquad (5)$$

by denoting $w_1$ as $w$. Out of 9 possible combinations of the sub-retention functions (i.e., BC, VG, and KO), this paper focuses on three models: a dual-BC (DB) model with BC type $S_1(h)$ and $S_2(h)$ equations, a dual-VG (DV) model with VG type $S_1(h)$ and $S_2(h)$ equations (this model is equivalent to the one proposed by Durner (1994)), and a KO$_1$BC$_2$ (KB) model with KO type $S_1(h)$ and BC type $S_2(h)$ equations.

We additionally use the CH (common head) assumption of Seki et al. (2022) in which $h_{b_i}$ for the BC model, $\alpha_i^{-1}$ for the VG model, and $h_{m_i}$ for the KO model have the same value (i.e., $H = h_{b_i} = \alpha_i^{-1} = h_{m_i}$). The CH assumption is very useful for soils showing a single inflection point in the water retention data (i.e., still having a unimodal pore-size distribution) but is ineffective for soils which have a clear bimodality of soil pore sizes (such as Andisols). Since the verification dataset used in this study does not include such soils, we use for this scenario the CH assumption. To summarize, this paper verifies dual-BC-CH (DBC), dual-VG-CH (DVC) and KO$_1$BC$_2$-CH (KBC) models as summarized in Table 2.

**Table 2**. Parameters in the selected water retention (WRF) and hydraulic conductivity (HCF) functions. Fixed parameters are shown within bracket, with $h_b$=2 cm and $h_0$=6.3x10$^6$ cm; other fixed values are shown in the table.

| Model name | Abbreviation | WRF parameters | HCF parameters |
|---|---|---|---|
| van Genuchten | VG | $\theta_s$, $\theta_r$, $H$, $n$, ($q$=1) | $K_s$, $p$, ($r$=2) |
| Modified VG | MVG | $\theta_s$, $\theta_r$, $H$, $n$, ($h_b$, $q$=1) | $K_s$, $p$, ($r$=1) |
| dual-BC-CH | DBC | $\theta_s$, $w$, $H$, $\lambda_1$, $\lambda_2$ | $K_s$, $p$, $q$, ($r$=1) |
| dual-VG-CH | DVC | $\theta_s$, $w$, $H$, $n_1$, $n_2$, ($q$=1) | $K_s$, $p$, $r$ |
| Modified DVC | MDVC | $\theta_s$, $w$, $H$, $n_1$, $n_2$, ($h_b$, $q$=1) | $K_s$, $p$, $r$ |
| KO$_1$BC$_2$-CH | KBC | $\theta_s$, $w$, $H$, $\sigma_1$, $\lambda_2$ | $K_s$, $p$, $q$, ($r$=1) |
| Modified KBC | MKBC | $\theta_s$, $w$, $H$, $n_1$, $n_2$, ($h_b$) | $K_s$, $p$, $q$, ($r$=1) |
| Peters | PE | $\theta_s$, $w$, $H$, $\sigma$, ($h_0$) | $K_s$, $p$, $a$, $\omega$ |
| Modified PE | MPE | $\theta_s$, $w$, $H$, $\sigma$, ($h_b$, $h_0$) | $K_s$, $p$, $a$, $\omega$ |

We also compared the VG (van Genuchten, 1980) and PE (Peters, 2013) models. The WRF of Peters (2013) is given by Eq. (5), in which $S_1(h)$ is a KO type subfunction as shown in Table 1 with $h_m$=$H$, while $S_2(h)$ is given by

$$S_2(h) = \begin{cases} \dfrac{L(h_0) - L(h)}{L(h_0) - L(H)} & (h > H) \\ 1 & (h \leq H) \end{cases} \quad (6)$$

where $L(h) = \log(1 + h/H)$. Note that $h_0$ is the pressure head where $\theta$ first becomes zero. The HCF of the PE model is expressed as

$$K_r(h) = (1-\omega) K_1(h) + \omega K_2(h) \quad (7)$$

where $K_1(h)$ is Mualem's equation (Eq. (1) with $q$=1, $r$=2) for $S_1(h)$, while $K_2(h)$ is a BC type function given by

$$K_2(h) = \begin{cases} \left(\dfrac{h}{H}\right)^a & (h > H) \\ 1 & (h \leq H) \end{cases} \quad (8)$$

where $K_s$ and $p$ in Eq. (1), $\omega$ in Eq. (7) and $a$ in Eq. (8) are HCF parameters of the PE model. When $n$<1.1 in the VG or DVC models, or $\sigma$>2 in the KBC or PE models, we used a modified formulation which introduces a hypothetical air-entry head near saturation ($h_b$=2 cm) as proposed by Vogel et al. (2000) and was adapted to the multimodal model by Seki et al.

(2022) to suppress extreme changes in the HCF curve near saturation. These modified models are denoted as MVG, MDVG, MKBC and MPE in Table 2.

**Reason for optimizing 2 HCF variables**

Although the original Burdine and Mualem models assume fixed values of the parameters $p$, $q$, $r$ in the general hydraulic conductivity function given by Eq. (1), it is possible to optimize the tortuosity factor, $p$, in both models. This section discusses why it may be advantageous to optimize two of the three parameters, notably ($p$, $q$) or ($p$, $r$) simultaneously for the dual-porosity models shown in Table 2.

As shown by Seki et al. (2022), the slope $a_i$ of a log $h$ – log $K$ plot of a multi-BC model in which $i$-th subcurve is predominant can be approximated by
$$a_i = (p + r)\lambda_i + qr \tag{9}$$
with $a_i$ having a positive value. By defining $\alpha = p+r$ and $\beta = qr$, the relationship between $\lambda_i$ and $a_i$ is given by a linear function $a_i = \alpha\lambda_i + \beta$, which indicates that a pair of parameters ($\alpha$, $\beta$) determines the shape of the HCF curve. For a single BC model, we only have one equation for the relationship between $a_1$ and $\lambda_1$, which implies that when $q$ and $r$ are fixed, $p$ can be optimized to fix the values of $a_1$ and $\lambda_1$. For the dual-BC model, on the other hand, one has two equations for the ($a_i$, $\lambda_i$) relationship, which shows that two free parameters of ($p$, $q$, $r$) are needed to solve the simultaneous equations. For the dual-modal function it is hence reasonable to optimize two parameters of ($p$, $q$, $r$) simultaneously.

However, making all three parameters ($p$, $q$, $r$) freely adjustable is not reasonable, even for multi-models with $k$=3 (the tri-model), because, as noted above, $\alpha$ and $\beta$ are the only free parameters such that we have only two degrees of freedom. When using the tri-model it is hence also reasonable to optimize only two of the parameters $p$, $q$, and $r$. Seki et al. (2022) showed that for any multimodal model, the fitted water retention parameters will produce very similar hydraulic conductivity curves as long as the same ($p$, $q$, $r$) parameter set is used. Therefore, the discussion of this section with multi-BC model can be generalized to any combinations of sub-retention functions, as will be verified with actual soil data later.

**Parameter optimization**

The WRF and HCF for the models in Table 2 were tested using selected $\theta(h)$ and $K(h)$ data sets from Mualem (1976) and the UNSODA unsaturated soil hydraulic database (Nemes et al., 1999). The WRF parameters in Table 2 were determined using a least-squares optimization method in which the mean squared error (MSE) between the estimated and measured $\theta$ values was minimized. The HCF parameters were optimized subsequently by minimizing the MSE errors between estimated and measured log($K$) values. For the HCF optimization we kept the WRF derived parameters constant and only adjusted the remaining HCF parameters. For the PE model we followed Peters (2013) and fixed $h_0$ at $6.3 \times 10^6$ cm. Although Peters (2013) assumed $a$=-1.5 to be a constant, in this study $a$ was also optimized to assess the flexibility of the model for multimodal data sets.

Except for $K_s$, Eq. (1) has 3 HCF parameters: $p$, $q$, and $r$. For the multimodal models we fixed one of these three parameters and optimized the remaining two parameters as explained in the last section. For the DBC and KBC models, we used $r$=1 and optimized $p$ and $q$. We selected $r$=1 since Kosugi (1999) showed that $r$=1 is ideal for a single KO model (Fig. 4 in his paper), in which case the HCF expression of multimodal model can be simplified to a sum of sub-functions (Seki et al., 2022). The same method was not used for the DVC model since $q$ is a DVC water retention parameter and cannot be used as an optimized parameter for the hydraulic conductivity once the WRF parameters are known. Since $q$=1 is widely used for the VG model, we also used this value for the DVC model and optimized $p$ and $r$. For the VG model we used Mualem's hydraulic conductivity model ($q$=1, $r$=2) and optimized $p$. For the PE model, *p*, *a*, *ω* were optimized as explained earlier. The optimized and fixed parameters are summarized in Table 2.

The HCF parameters were optimized in two steps. At first, multiple initial conditions, notably (1, 2, 4, 6) for *p* and (0.5, 1, 2) for *q* or *r* (12 combinations in total) were used and the parameters optimized with a relatively loose convergence criterion to facilitate rapid calculations. The fitted parameter set with the least MSE was used next as initial condition for a second optimization step using much stricter convergence criteria to obtain more accurate parameter values. The software used for the WRF and HCF parameter optimizations is published as a Python library named unsatfit. On the website of unsatfit (i.e., https://sekika.github.io/unsatfit/), sample codes are provided for determining the WRF and HCF parameters by using the same optimization strategy as used here and also for drawing similar figures as published in this paper. A relatively simple web interface (SWRC Fit) for optimizing water retention parameters of the multimodal models is also available (Seki, 2007).

**RESULTS AND DISCUSSION**

Below are results obtained with the various models. We first present results obtained for Gilat loam since this soil was also used by Peters (2013) as an example of hydraulic properties covering a wide range of pressure heads. A detailed sensitivity analysis of the parameter optimization of Gilat loam is presented next. We subsequently show applications of the models to many other soils.

**Gilat loam**

Figure 1 shows fitted water retention and hydraulic conductivity curves of Gilat loam (Mualem, 1976), which was analyzed also (as soil 4) by Peters (2013). Fixed values of $\theta_s$=0.44 and $K_s$=1.37x10$^{-4}$ cm/s were used consistent with the optimization procedure adopted by Peters (2013). The DBC, DVC, KBC and PE models all fitted the WRF and HCF data over the whole range of pressure heads, including the adsorption moisture range, nearly perfectly. This result confirms that the general HCF is effective for a wide range of pressure heads.

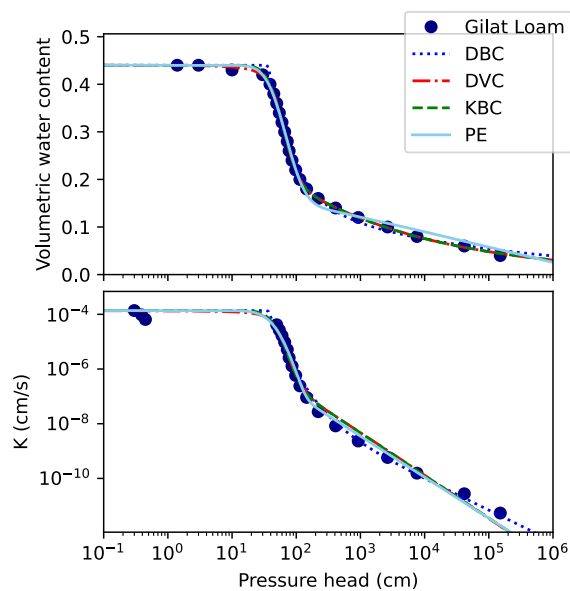

**Fig. 1.** Water retention and hydraulic conductivity curves for Gilat loam fitted with the DBC, DVC, KBC and PE models.

**Sensitivity analysis**

Fig. 2 shows a contour plot of RMSE (the root mean squared error) of the estimated $\log_{10}(K)$ for the KBC ($r=1$) optimization of Gilat loam, along with the optimized values of ($p$, $q$). It is clear that the original Burdine parameter set ($p$, $q$) = (2, 2) did not give the lowest RMSE, while optimizing $p$ with a fixed $q=2$ also did not give the lowest RMSE.

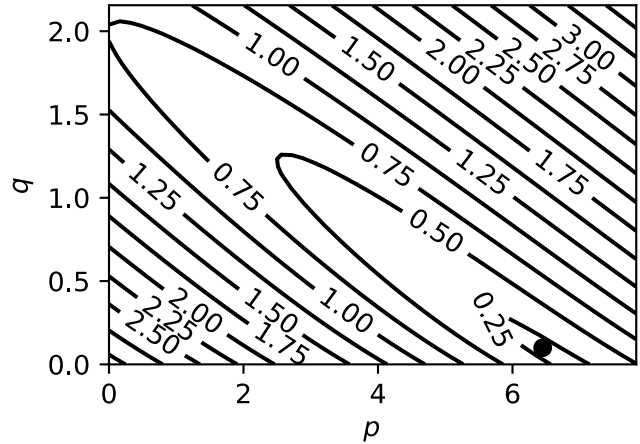

**Fig. 2.** Contour lines of RMSE of $\log_{10}(K)$ using various $p$ and $q$ values, and the optimized parameter (closed circle) obtained with the KBC model ($r=1$) for Gilat loam.

Using Burdine's setting of ($q$, $r$) = (2, 1), the HCF curve did not match the entire range of pressure head, as shown in Fig. 3a1; the curve in the low pressure head range ($h<100$ cm) matched the data well when $p=4$, but the calculated curve at higher pressure heads ($h>100$ cm) did not match the data. However, the measured and calculated curves matched over the whole range of pressure heads when ($p$, $q$, $r$) = (6.4, 0.1, 1) (Fig. 3a3). For the DVC optimization ($q=1$), similar curves can be drawn as shown in Fig. 3b. For Mualem's setting of ($q$, $r$)=(1, 2) (Fig. 3b1), the HCF curve did not match the observed data over the whole range of pressure heads, while the curve matched the data for all pressure heads when ($p$, $q$, $r$) = (7.3, 1, 0.1) (Fig. 3b3). As noted earlier from Eq. (9) that the ($p+r$, $qr$) pair is a critical parameter set, Fig. 3b was drawn so that the ($p+r$, $qr$) values corresponded to those for Fig. 3a. As a result, similar curves were drawn for each corresponding KBC and DVC optimization, especially for the second (a2 and b2) and the third (a3 and b3) plots.

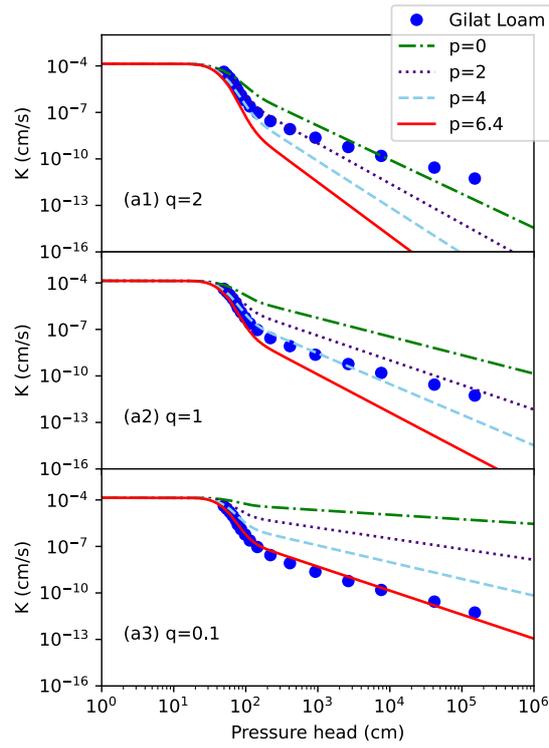

**Fig. 3a.** Hydraulic conductivity curves for Gilat loam as obtained with the KBC model using a fixed *r*=1 and changing *p*, for (a1) *q*=2 (a2) *q*=1 (a3) *q*=0.1.

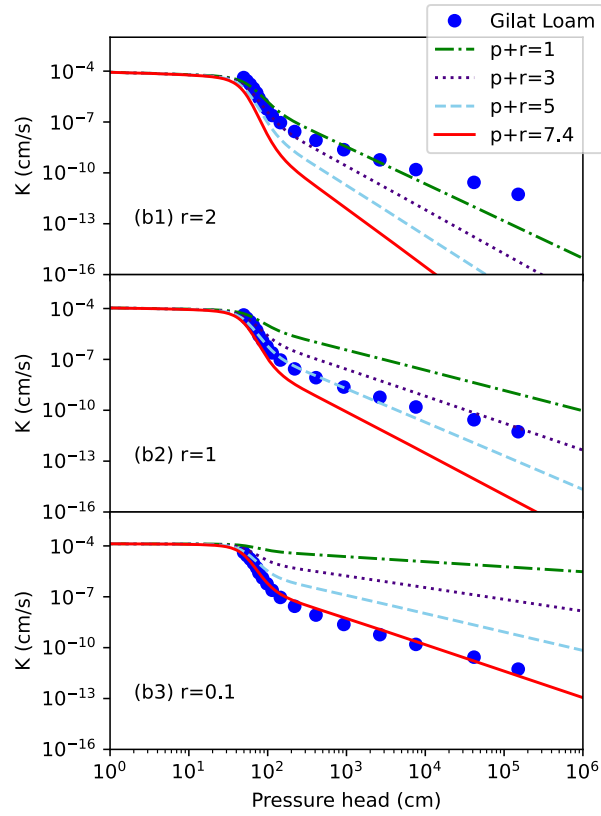

**Fig. 3b.** Hydraulic conductivity curves of Gilat loam as obtained with the DVC model using a fixed $q=1$ and changing $p$, for (b1) $r=2$ (b2) $r=1$ (b3) $r=0.1$.

Fig. 3 demonstrates that the slope of the log $h$ – log $K$ curve in the low and high pressure head ranges can be controlled with two parameters $p$ and $q$ and a constant $r=1$ (Fig. 3a), or alternatively with $p$ and $r$ and a constant $q=1$ (Fig. 3b), as emphasized in the Methodology section. For Gilat loam, the dual-BC-CH fitting resulted in $\lambda_1=1.15$ and $\lambda_2=0.14$, and $(p, q, r)$ = (6.4, 0.1, 1), and hence $a_1=8.6$ and $a_2=1.14$ when using Eq. (9). That equation hence indicates that points $(\lambda_i, a_i)$ are on the straight line $a=(p+r)\lambda+qr$ in a $\lambda$-$a$ plane, as shown in Fig. 4 for Gilat loam. The optimized values $(\lambda_1, a_1) = (1.15, 8.6)$ and $(\lambda_2, a_2) = (0.14, 1.14)$ are both on the red straight line $a=(p+r)\lambda+qr$ with the optimized $(p, q, r) = (6.4, 0.1, 1)$. In other words, for a soil with WRF parameters $(\lambda_1, \lambda_2)$ and corresponding slopes of the log $h$ – log $K$ curve $(a_1, a_2)$, the parameters $(p, q, r)$ are optimized to values corresponding to the red line in this Fig. 4. If we use Burdine's parameters of $q=2$ and $r=1$ or Mualem's parameters of $q=1$ and $r=2$, the line is expressed as $a=(p+r)\lambda+2$ and fixed to the (0, 2) point, and hence cannot lie on $(\lambda_1, a_1)$ and $(\lambda_2, a_2)$ simultaneously. For this condition we can only optimize to the capillary point $(\lambda_1, a_1)$, where the blue dotted line $(a=5.7\lambda+2)$ is obtained. This corresponds to $p=4.7, q=2, r=1$ or $p=3.7, q=1, r=2$, which causes the slope of the HCF curve to be steeper than the measured values in the high pressure head range as demonstrated in Figs. 3a1 and 3b1.

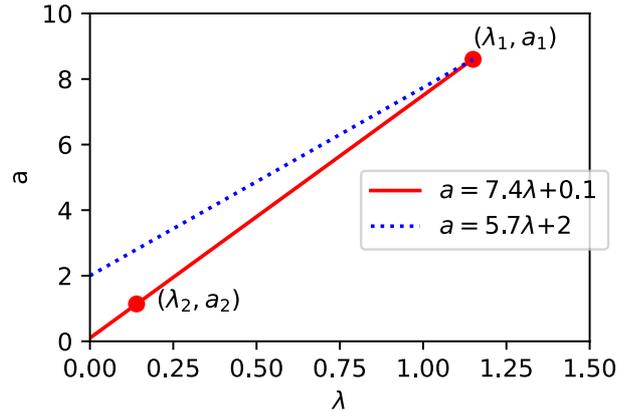

**Fig. 4.** Illustration that the ($\lambda_i$, $a_i$) points are on the same line $a=(p+r)\lambda+qr$ (red straight line) in a $\lambda$-$a$ plane for Gilat loam. The blue-dotted line connects (0, 2) and ($\lambda_1$, $a_1$), which expresses the condition that the product $q\,r$ equals 2 for the Mualem ($q=1$, $r=2$) or Burdine ($q=2$, $r=1$) models.

As discussed earlier, Eq. (9) suggests that different parameter sets for ($p$, $q$, $r$), but with the same ($p+r$, $qr$), all will produce similar HCF curves for all multimodal models. This is verified in Fig. 5. As we obtained ($p$, $q$, $r$) = (6.4, 0.1, 1) for the optimized ($p$, $q$) with fixed $r$ for the KBC model, different parameter sets with the same ($p+r$, $qr$) = (7.4, 0.1) were chosen. We found that all curves were almost identical, which justifies that only two parameters need to be selected for the optimizations. Since the ($p+r$, $qr$) pair has two degrees of freedom, optimizing three parameters ($p$, $q$, $r$) simultaneously will cause overparameterization.

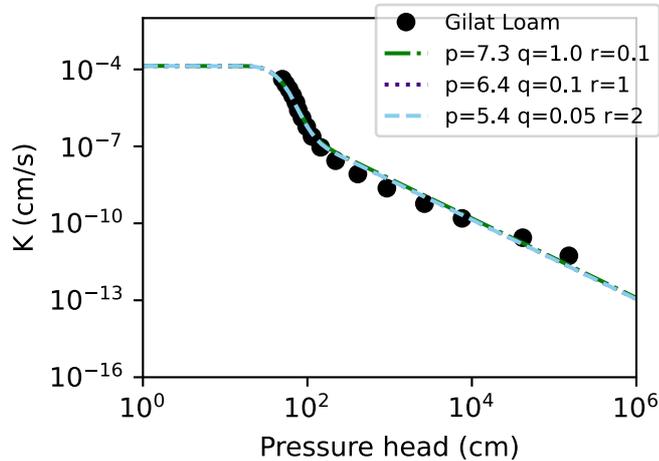

**Fig. 5.** Hydraulic conductivity curves of Gilat loam as obtained with the KBC model using different fixed values of ($p$, $q$, $r$), where ($p+r$, $qr$) = (7.4, 0.1).

**Other soils**

Fig. 6 shows fitted curves for 20 soils taken from the UNSODA database: 2 clays, 1 clay loam, 3 loams, 2 loamy sands, 6 sands, 1 silty clay, and 5 silty loams. The optimized parameters are listed in the Appendix.

For the examined datasets, all of the multimodal models represented the overall shape of WRF and HCF very well. For clay soils C2360 and C2362 and loam soil L4592 (Fig. 6a), the VG and MVG model fitted the WRF and HCF data well over a wide range of pressure heads, which shows that in this case there is no need to use multimodal models. For CL3033 in Fig. 6a, L4780 in Fig. 6b, S3182 in Fig. 6c, S4661, SiL3370 in Fig. 6d and SiL4673 in Fig. 6e, the VG and MVG models did not express the change in the slope well between the low (< 100 cm) and high pressure heads, while also underestimating the hydraulic conductivity at high pressure heads. The multimodal models (DBC, DVC, MDVC, KBC, MKBC) and the PE and MPE models on the other hand closely matched the water retention and hydraulic conductivity data within the measurement range. The DBC formulation furthermore was suitable for soils with a distinct air-entry head, such as L4770 in Fig. 6b, but for soils without a distinct air-entry head (such as the silty clay (SiC) and silty loam (SiL) soils in Figs. 6d and 6e), the DVC and KBC performed better than DBC.

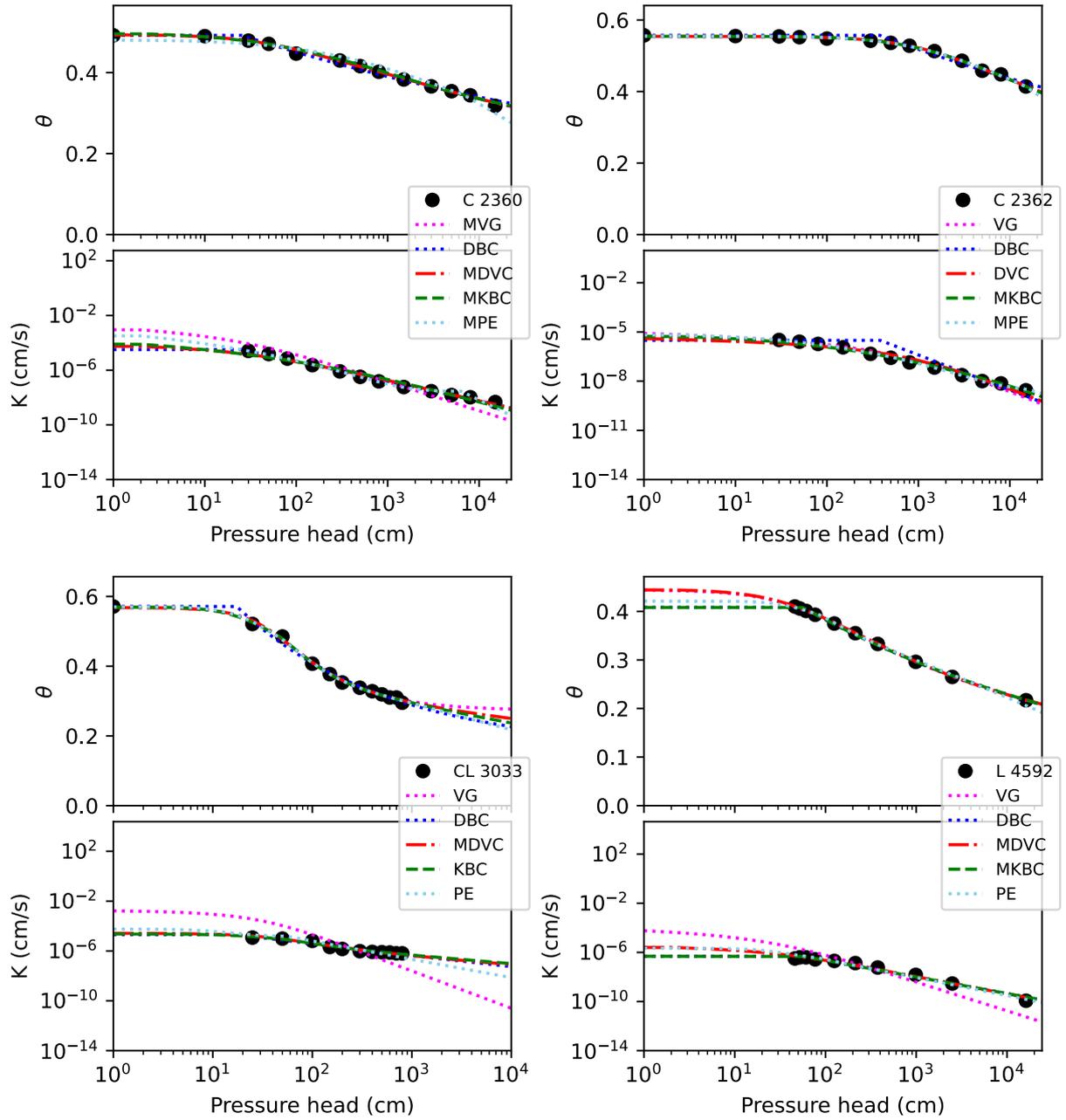

**Fig. 6a**. Measured water retention and hydraulic conductivity data and fitted curves for soil samples identified with the abbreviated name of soil texture and the UNSODA ID. Data for $h=0$ are shown at $h=1$ cm.

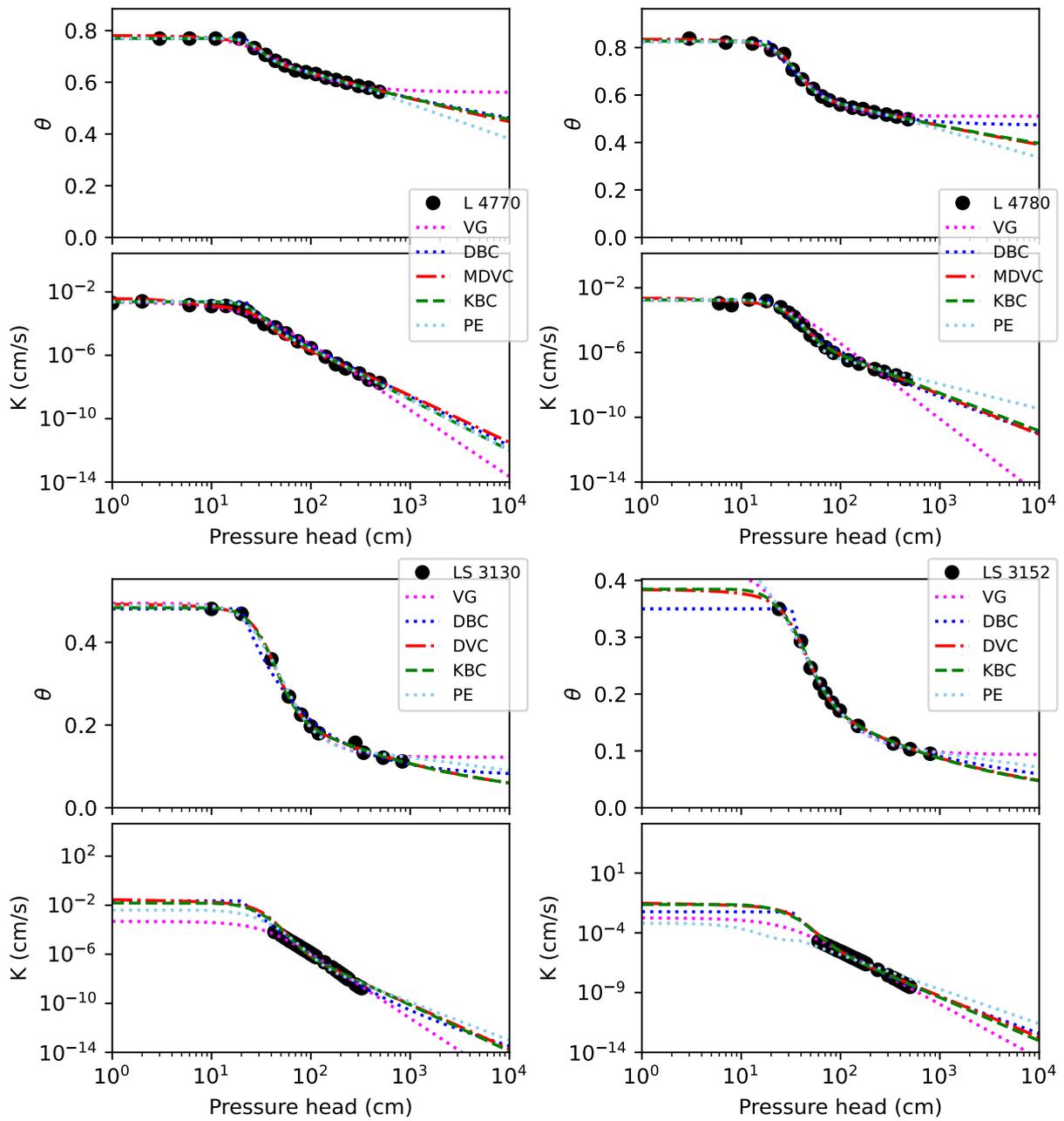

**Fig. 6b**. Measured water retention and hydraulic conductivity data and fitted curves for soil samples identified with the abbreviated name of soil texture and the UNSODA ID. Data for *h*=0 are shown at *h*=1 cm.

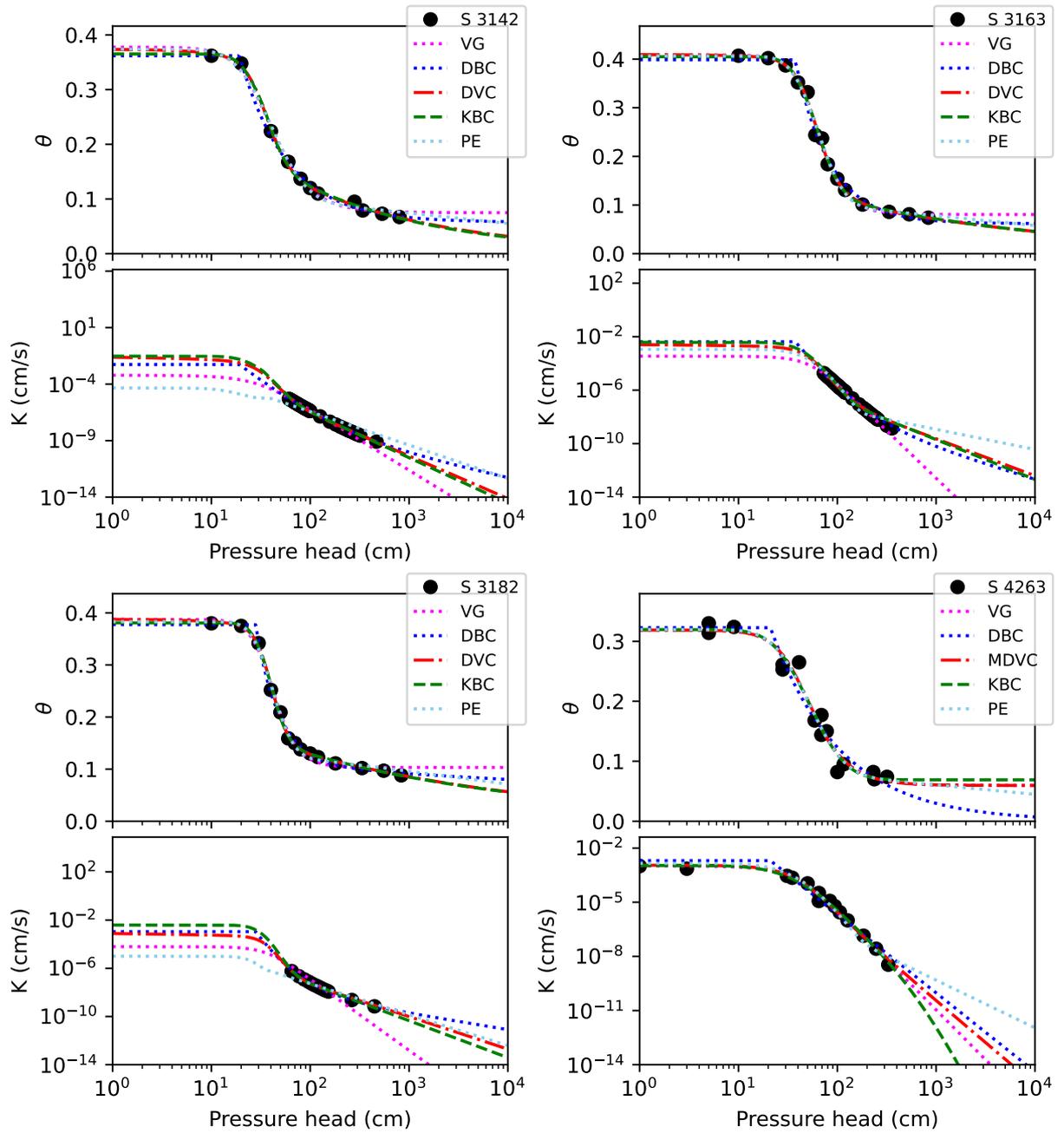

**Fig. 6c**. Measured water retention and hydraulic conductivity data and fitted curves for soil samples identified with the abbreviated name of soil texture and the UNSODA ID. Data for *h*=0 are shown at *h*=1 cm.

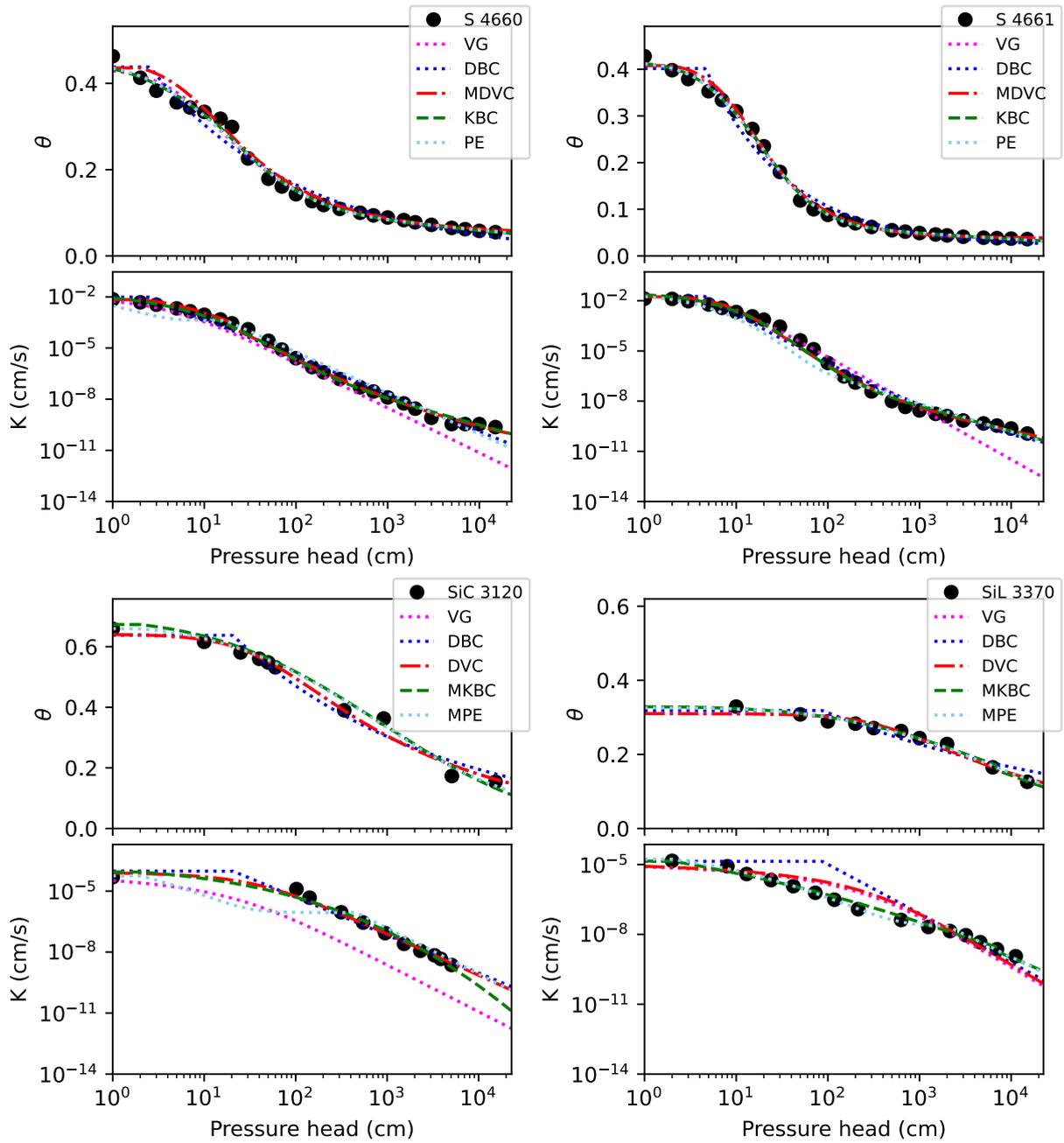

**Fig. 6d**. Measured water retention and hydraulic conductivity data and fitted curves for soil samples identified with the abbreviated name of soil texture and the UNSODA ID. Data for *h*=0 are shown at *h*=1 cm.

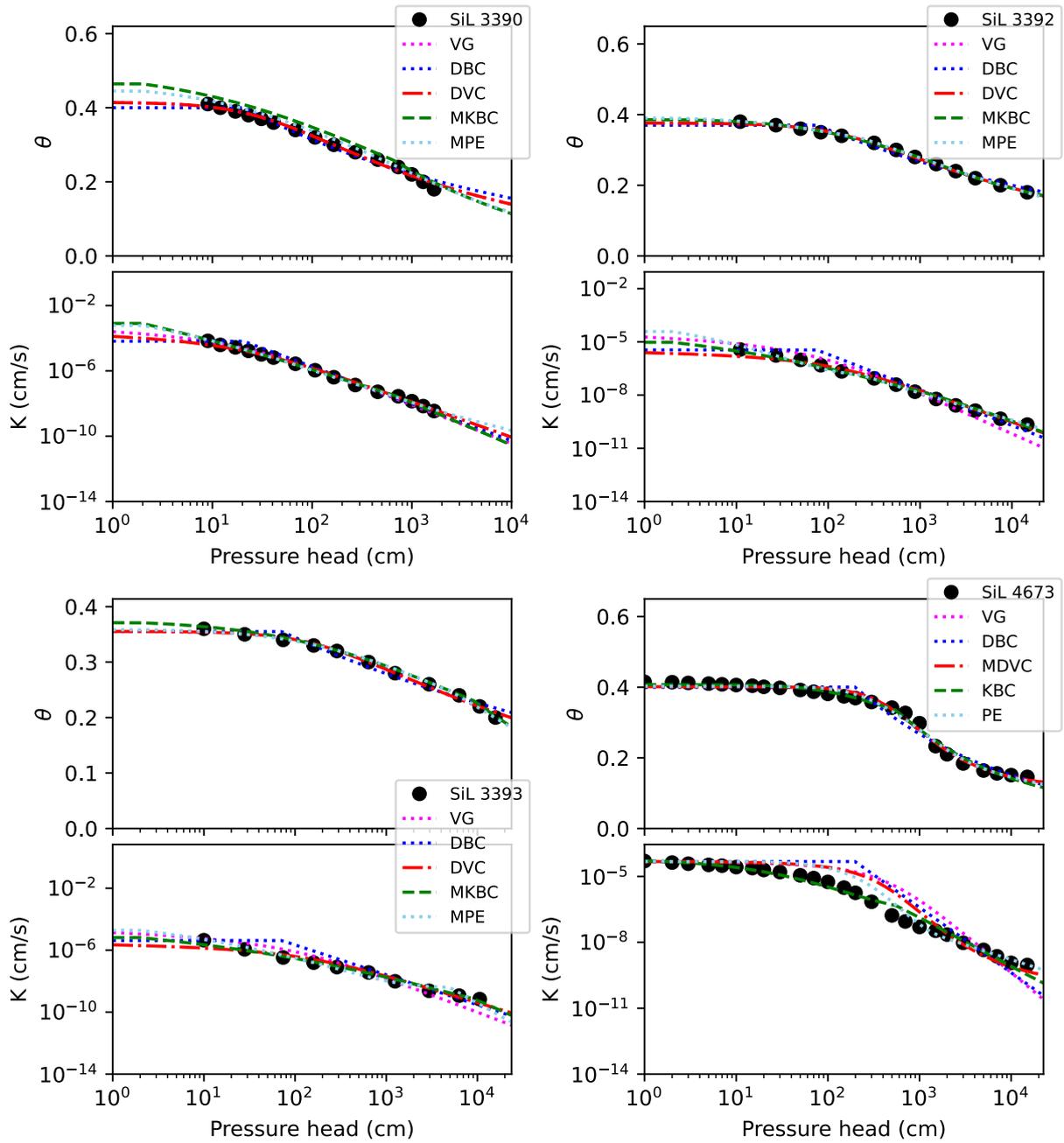

**Fig. 6e**. Measured water retention and hydraulic conductivity data and fitted curves for soil samples identified with the abbreviated name of soil texture and the UNSODA ID. Data for $h=0$ are shown at $h=1$ cm.

Table 3 summarizes the coefficients of determination ($R^2$) for the fitted curves of Fig. 6. The average $R^2$ for $\theta$ was highest (0.992) for the (modified) DVC model, while $R^2$ for log $K$ was highest (0.985) for the (modified) KBC. DBC had the highest $R^2$ for log $K$ for 5 soils, even though the average value was not very high (0.938). $R^2$ for log $K$ was higher for the (modified) KBC compared to the (modified) PE for 13 soils, and higher for the (modified)

DVC relative to the (modified) PE model for 11 soils. Although the DVC and KBC models had fewer parameters than PE (Table 2), they can be used as an alternative to the PE model with similar or better fitting performance for many soils.

**Table 3**. Coefficient of determination ($R^2$) of $\theta$ and log $K$ for the models fitted in this study. The best fit (largest $R^2$) values for a particular soil sample are shown in bold. Total best-fit numbers are shown in the bottom row. Soil samples are identified with their abbreviated soil texture name and the UNSODA ID.

| Sample | VG, MVG | | DBC | | DVC, MDVC | | KBC, MKBC | | PE, MPE | |
|---|---|---|---|---|---|---|---|---|---|---|
| | $\theta$ | log $K$ | $\theta$ | log $K$ | $\theta$ | log $K$ | $\theta$ | log $K$ | $\theta$ | log $K$ |
| C2360 | **0.9958** | 0.8330 | 0.9887 | **0.9931** | 0.9958 | 0.9823 | 0.9936 | 0.9710 | 0.9607 | 0.9845 |
| C2362 | **0.9968** | 0.9650 | 0.9727 | 0.8087 | 0.9968 | 0.9689 | 0.9967 | 0.9897 | 0.9965 | **0.9968** |
| CL3033 | 0.9968 | -2.030 | 0.9918 | 0.9416 | 0.9968 | 0.9536 | **0.9983** | 0.9639 | 0.9976 | 0.8168 |
| L4592 | **0.9999** | 0.6115 | 0.9988 | 0.9783 | 0.9997 | **0.9831** | 0.9988 | 0.9783 | 0.9955 | 0.9767 |
| L4770 | 0.9868 | 0.9867 | **0.9989** | 0.9771 | 0.9954 | 0.9899 | 0.9984 | 0.9893 | 0.9964 | **0.9902** |
| L4780 | 0.9943 | 0.9042 | 0.9907 | 0.9852 | **0.9978** | **0.9949** | 0.9970 | 0.9945 | 0.9968 | 0.9917 |
| LS3130 | 0.9955 | 0.9964 | 0.9900 | **0.9935** | 0.9986 | 0.9554 | **0.9986** | 0.9564 | 0.9956 | 0.9711 |
| LS3152 | 0.9971 | **0.9961** | **0.9998** | 0.9817 | 0.9991 | 0.9748 | 0.9992 | 0.9789 | 0.9967 | 0.9041 |
| S3142 | 0.9946 | 0.9620 | 0.9973 | **0.9998** | 0.9981 | 0.9966 | 0.9980 | 0.9968 | 0.9936 | 0.8869 |
| S3163 | 0.9942 | 0.9863 | 0.9875 | **0.9958** | 0.9948 | 0.9813 | 0.9942 | 0.9765 | 0.9938 | 0.9700 |
| S3182 | 0.9928 | 0.4823 | 0.9972 | **0.9969** | 0.9979 | 0.9889 | 0.9977 | 0.9912 | 0.9953 | 0.9033 |
| S4263 | 0.9683 | 0.9925 | 0.9539 | 0.9656 | 0.9683 | 0.9889 | **0.9703** | **0.9927** | 0.9699 | 0.9776 |
| S4660 | 0.9890 | 0.9191 | 0.9747 | 0.9877 | 0.9813 | **0.9959** | 0.9936 | 0.9956 | 0.9926 | 0.9760 |
| S4661 | 0.9966 | 0.9322 | 0.9842 | 0.9858 | 0.9949 | 0.9943 | **0.9981** | **0.9944** | 0.9979 | 0.9728 |
| SiC3120 | 0.9842 | -0.078 | 0.9703 | 0.9806 | 0.9842 | **0.9869** | 0.9842 | 0.9833 | **0.9883** | 0.8924 |
| SiL3370 | 0.9695 | 0.9029 | 0.9163 | 0.6737 | 0.9695 | 0.8927 | **0.9882** | 0.9917 | 0.9871 | **0.9970** |
| SiL3390 | **0.9920** | 0.9923 | 0.9799 | 0.9716 | 0.9920 | 0.9949 | 0.9167 | 0.9977 | 0.9738 | **0.9981** |
| SiL3392 | 0.9978 | 0.9145 | 0.9817 | 0.9196 | 0.9978 | 0.9852 | **0.9993** | 0.9928 | 0.9989 | **0.9937** |
| SiL3393 | 0.9925 | 0.8679 | 0.9732 | 0.8229 | 0.9925 | 0.9650 | 0.9947 | **0.9878** | **0.9955** | 0.9746 |
| SiL4673 | 0.9885 | 0.8143 | 0.9715 | 0.8059 | 0.9885 | 0.8948 | **0.9914** | **0.9812** | 0.9887 | 0.9443 |
| Average | 0.9911 | 0.6976 | 0.9810 | 0.9383 | **0.9920** | 0.9734 | 0.9903 | **0.9852** | 0.9905 | 0.9559 |
| Total best | 4 | 1 | 2 | 5 | 4 | 4 | 8 | 5 | 2 | 5 |

Our results confirm that the water retention and hydraulic conductivity properties can be expressed with a multimodal model in the same way as the PE model, by expressing capillary water retention with the first subfunction and adsorptive water with the second subfunction, while the hydraulic conductivity characteristics can be described using two parameters such as ($p$, $q$) or ($p$, $r$).

## CONCLUSIONS

Multimodal water retention (WRF) and hydraulic conductivity (HCF) equations proposed by Seki et al. (2022) were shown to be effective with different types of soils. Our study confirms that the selected models are very flexible in terms of fitting observed WRF and HCF curves of various soils, especially when optimizing two parameters in the general conductivity equation, notably ($p$, $q$) or ($p$, $r$), in addition to $K_s$. Among the verified models, the KO$_1$BC$_2$-CH (KBC) model with $r$=1 and the optimized ($p$, $q$) parameters performed best, with the dual-VG-CH (DVC) model with $q$=1 also performing well and being compatible with the original Durner (1994) multimodal formulation. The latter model showed a good fitting performance when ($p$, $r$) were optimized simultaneously. We conclude that the proposed model formulations should be very useful for practical applications, while mathematically being relatively simple and consistent.

**ABBREVIATIONS**

| | |
|---|---|
| BC | Brooks and Corey |
| C | Clay |
| CH | Common head |
| CL | Clay loam |
| DB | dual-BC |
| DBC | dual-BC-CH |
| DV | dual-VG |
| DVC | dual-VG-CH |
| HCF | Hydraulic conductivity function |
| KB | $KO_1BC_2$ |
| KBC | $KO_1BC_2$-CH |
| KO | Kosugi |
| L | Loam |
| LS | Loamy sand |
| MDVC | Modified dual-VG-CH |
| MKBC | Modified $KO_1BC_2$-CH |
| MPE | Modified Peters |
| MSE | Mean squared error |
| MVG | Modified van Genuchten |
| PE | Peters |
| RMSE | Root mean squared error |
| S | Sand |
| SiC | Silty clay |
| SiL | Silty loam |
| UNSODA | Unsaturated soil hydraulic database |
| VG | van Genuchten |
| WRF | Water retention function |

Appendix. Optimized parameters of the examples shown in Fig. 6.

Clay soil C 2360

| Model | $\theta_s$ | $\theta_r$ | $w$ | $H$ (cm) | $N_1$[a] | $N_2$[a] | $K_s$ (cm/s) | $p$ | $qr$ or $a$ | $\omega$ |
|---|---|---|---|---|---|---|---|---|---|---|
| MVG | 0.493 | 0.000 | | 47.9 | 1.07 | | 8.99.E-4 | -0.18 | 2[b] | |
| MKBC | 0.496 | 0[b] | 0.438 | 139.8 | 2.95 | 0.000 | 7.93.E-5 | 6.00 | 1.00 | |
| DBC | 0.492 | 0[b] | 0.788 | 25.3 | 0.079 | 0.011 | 3.19.E-5 | 6.00 | 1.00 | |
| MDVC | 0.493 | 0[b] | 0.772 | 47.9 | 1.07 | 1.07 | 5.53.E-5 | 6.00 | 1.00 | |
| MPE | 0.480 | 0[b] | 0.555 | 6379.8 | 3.00 | | 3.22.E-4 | 5.79 | -3.25 | 1.22.E-5 |

Clay soil C 2362

| Model | $\theta_s$ | $\theta_r$ | $w$ | $H$ (cm) | $N_1$[a] | $N_2$[a] | $K_s$ (cm/s) | $p$ | $qr$ or $a$ | $\omega$ |
|---|---|---|---|---|---|---|---|---|---|---|
| VG | 0.554 | 0.000 | | 1215.8 | 1.11 | | 2.69.E-5 | 0.00 | 2[b] | |
| MKBC | 0.554 | 0[b] | 0.397 | 8000.0 | 2.00 | 0.0148 | 5.38.E-6 | 0.42 | 1.52 | |
| DBC | 0.557 | 0[b] | 0.747 | 383.1 | 0.093 | 0.026 | 3.08.E-6 | 8.74 | 1.40 | |
| DVC | 0.554 | 0[b] | 0.773 | 1215.8 | 1.11 | 1.11 | 9.13.E-6 | 5.36 | 1.41 | |
| MPE | 0.555 | 0[b] | 0.510 | 14999.8 | 2.30 | | 6.57.E-6 | 1.00 | -1.50 | 3.76.E-4 |

Clay loam soil CL 3033

| Model | $\theta_s$ | $\theta_r$ | $w$ | $H$ (cm) | $N_1$[a] | $N_2$[a] | $K_s$ (cm/s) | $p$ | $qr$ or $a$ | $\omega$ |
|---|---|---|---|---|---|---|---|---|---|---|
| VG | 0.569 | 0.272 | | 41.6 | 1.77 | | 1.88.E-3 | -0.70 | 2[b] | |
| KBC | 0.569 | 0[b] | 0.313 | 49.9 | 1.01 | 0.0944 | 2.29.E-5 | 1.00 | 0.50 | |
| DBC | 0.571 | 0[b] | 0.719 | 17.4 | 0.29 | 0.000 | 2.07.E-5 | 1.00 | 0.50 | |
| MDVC | 0.568 | 0[b] | 0.384 | 43.2 | 2.02 | 1.06 | 2.63.E-5 | 2.02 | 0.50 | |
| PE | 0.570 | 0[b] | 0.350 | 53.5 | 0.959 | | 5.69.E-5 | 9.57 | -1.47 | 2.77.E-1 |

Loam soil L 4592

| Model | $\theta_s$ | $\theta_r$ | w | H (cm) | $N_1$[a)] | $N_2$[a)] | $K_s$ (cm/s) | p | qr or a | $\omega$ |
|---|---|---|---|---|---|---|---|---|---|---|
| VG | 0.444 | 0.092 | | 46.4 | 1.18 | | 2.39.E-4 | -0.01 | 2 [b)] | |
| MKBC | 0.408 | 0[b)] | 0.000 | 56.2 | 2.50 | 0.111 | 4.83.E-7 | 0.93 | 1.13 | |
| DBC | 0.408 | 0[b)] | 0.500 | 56.2 | 0.111 | 0.111 | 4.83.E-7 | 0.93 | 1.13 | |
| MDVC | 0.444 | 0[b)] | 0.794 | 46.4 | 1.18 | 1.00 | 2.55.E-6 | 2.01 | 1.00 | |
| PE | 0.421 | 0[b)] | 0.140 | 88.9 | 0.865 | | 2.17.E-6 | 6.15 | -1.46 | 1.38.E-1 |

Loam soil L 4770

| Model | $\theta_s$ | $\theta_r$ | w | H (cm) | $N_1$[a)] | $N_2$[a)] | $K_s$ (cm/s) | p | qr or a | $\omega$ |
|---|---|---|---|---|---|---|---|---|---|---|
| VG | 0.778 | 0.561 | | 32.8 | 1.98 | | 2.41.E-3 | 0.25 | 2 [b)] | |
| KBC | 0.770 | 0[b)] | 0.100 | 29.0 | 0.322 | 0.0713 | 2.41.E-3 | 9.32 | 2.47 | |
| DBC | 0.769 | 0[b)] | 0.090 | 21.9 | 2.22 | 0.0679 | 2.41.E-3 | 9.25 | 2.42 | |
| MDVC | 0.780 | 0[b)] | 0.098 | 33.9 | 5.42 | 1.08 | 3.69.E-3 | 3.93 | 2.46 | |
| PE | 0.770 | 0[b)] | 0.111 | 29.9 | 0.310 | | 2.41.E-3 | 8.64 | -3.22 | 5.33.E-2 |

Loam soil L 4780

| Model | $\theta_s$ | $\theta_r$ | w | H (cm) | $N_1$[a)] | $N_2$[a)] | $K_s$ (cm/s) | p | qr or a | $\omega$ |
|---|---|---|---|---|---|---|---|---|---|---|
| VG | 0.834 | 0.511 | | 32.6 | 2.75 | | 1.78.E-3 | -0.47 | 2[b)] | |
| KBC | 0.827 | 0[b)] | 0.264 | 32.9 | 0.462 | 0.0749 | 1.78.E-3 | 9.43 | 1.55 | |
| DBC | 0.825 | 0[b)] | 0.428 | 19.0 | 0.789 | 0.000 | 1.78.E-3 | 10.00 | 1.36 | |
| MDVC | 0.835 | 0[b)] | 0.256 | 33.3 | 4.27 | 1.08 | 2.22.E-3 | 4.00 | 2.00 | |
| PE | 0.826 | 0[b)] | 0.276 | 33.8 | 0.446 | | 1.78.E-3 | 1.00 | -1.52 | 1.11.E-3 |

Loamy sand soil LS 3130

| Model | $\theta_s$ | $\theta_r$ | w | H (cm) | $N_1$[a)] | $N_2$[a)] | $K_s$ (cm/s) | p | qr or a | $\omega$ |
|---|---|---|---|---|---|---|---|---|---|---|

| Model | $\theta_s$ | $\theta_r$ | $w$ | $H$ (cm) | $N_1$[a] | $N_2$[a] | $K_s$ (cm/s) | $p$ | $qr$ or $a$ | $\omega$ |
|---|---|---|---|---|---|---|---|---|---|---|
| VG | 0.495 | 0.122 | | 38.8 | 2.64 | | 4.88.E-4 | 0.01 | 2[b] | |
| KBC | 0.484 | 0[b] | 0.509 | 40.5 | 0.494 | 0.250 | 1.50.E-2 | 6.01 | 2.01 | |
| DBC | 0.481 | 0[b] | 0.840 | 19.8 | 0.680 | 0.000 | 2.24.E-2 | 8.14 | 1.39 | |
| DVC | 0.493 | 0[b] | 0.515 | 40.3 | 3.85 | 1.25 | 4.19.E-2 | 5.42 | 1.84 | |
| PE | 0.490 | 0[b] | 0.684 | 48.1 | 0.656 | | 4.08.E-3 | 2.35 | -3.11 | 3.91.E-4 |

Loamy sand soil LS 3152

| Model | $\theta_s$ | $\theta_r$ | $w$ | $H$ (cm) | $N_1$[a] | $N_2$[a] | $K_s$ (cm/s) | $p$ | $qr$ or $a$ | $\omega$ |
|---|---|---|---|---|---|---|---|---|---|---|
| VG | 0.418 | 0.094 | | 32.5 | 2.28 | | 1.81.E-3 | -0.02 | 2[b] | |
| KBC | 0.385 | 0[b] | 0.469 | 39.0 | 0.557 | 0.264 | 2.35.E-2 | 5.66 | 1.87 | |
| DBC | 0.350 | 0[b] | 0.534 | 32.1 | 1.46 | 0.176 | 5.77.E-3 | 5.75 | 1.80 | |
| DVC | 0.384 | 0[b] | 0.476 | 41.0 | 3.62 | 1.26 | 4.23.E-2 | 5.58 | 1.60 | |
| PE | 0.450 | 0[b] | 0.722 | 38.8 | 0.914 | | 6.36.E-4 | 9.10 | -2.88 | 3.46.E-2 |

Sand soil S 3142

| Model | $\theta_s$ | $\theta_r$ | $w$ | $H$ (cm) | $N_1$[a] | $N_2$[a] | $K_s$ (cm/s) | $p$ | $qr$ or $a$ | $\omega$ |
|---|---|---|---|---|---|---|---|---|---|---|
| VG | 0.378 | 0.075 | | 32.2 | 2.70 | | 6.13.E-4 | -0.03 | 2[b] | |
| KBC | 0.365 | 0[b] | 0.537 | 33.8 | 0.430 | 0.303 | 2.98.E-2 | 5.86 | 1.94 | |
| DBC | 0.362 | 0[b] | 0.841 | 19.2 | 0.900 | 0.000 | 5.41.E-3 | 6.03 | 1.03 | |
| DVC | 0.374 | 0[b] | 0.556 | 33.8 | 4.09 | 1.29 | 3.56.E-2 | 2.90 | 2.28 | |
| PE | 0.374 | 0[b] | 0.742 | 40.1 | 0.658 | | 4.60.E-5 | 8.87 | -2.96 | 1.19.E-1 |

Sand soil S 3163

| Model | $\theta_s$ | $\theta_r$ | $w$ | $H$ (cm) | $N_1$[a] | $N_2$[a] | $K_s$ (cm/s) | $p$ | $qr$ or $a$ | $\omega$ |
|---|---|---|---|---|---|---|---|---|---|---|
| VG | 0.409 | 0.081 | | 57.0 | 3.58 | | 3.47.E-4 | -0.02 | 2[b] | |
| KBC | 0.405 | 0[b] | 0.695 | 60.7 | 0.446 | 0.195 | 3.78.E-3 | 4.00 | 2.00 | |

| Model | $\theta_s$ | $\theta_r$ | $w$ | $H$ (cm) | $N_1$[a)] | $N_2$[a)] | $K_s$ (cm/s) | $p$ | $qr$ or $a$ | $\omega$ |
|---|---|---|---|---|---|---|---|---|---|---|
| DBC | 0.399 | 0[b)] | 0.845 | 37.5 | 1.23 | 0.000 | 4.20.E-3 | 6.01 | 1.01 | |
| DVC | 0.409 | 0[b)] | 0.697 | 57.0 | 4.12 | 1.19 | 3.41.E-3 | 2.03 | 2.01 | |
| PE | 0.407 | 0[b)] | 0.761 | 63.7 | 0.487 | | 1.14.E-3 | 1.00 | -1.50 | 6.43.E-5 |

Sand soil S 3182

| Model | $\theta_s$ | $\theta_r$ | $w$ | $H$ (cm) | $N_1$[a)] | $N_2$[a)] | $K_s$ (cm/s) | $p$ | $qr$ or $a$ | $\omega$ |
|---|---|---|---|---|---|---|---|---|---|---|
| VG | 0.388 | 0.103 | | 38.5 | 4.01 | | 6.27.E-5 | -0.70 | 2[b)] | |
| KBC | 0.381 | 0[b)] | 0.601 | 39.5 | 0.327 | 0.181 | 3.87.E-3 | 5.79 | 1.84 | |
| DBC | 0.377 | 0[b)] | 0.707 | 28.0 | 1.85 | 0.0546 | 1.10.E-3 | 5.95 | 0.98 | |
| DVC | 0.388 | 0[b)] | 0.602 | 38.7 | 5.35 | 1.18 | 1.22.E-3 | 2.00 | 2.00 | |
| PE | 0.383 | 0[b)] | 0.677 | 41.5 | 0.388 | | 1.01.E-5 | 8.29 | -2.57 | 4.97.E-2 |

Sand soil S 4263

| Model | $\theta_s$ | $\theta_r$ | $w$ | $H$ (cm) | $N_1$[a)] | $N_2$[a)] | $K_s$ (cm/s) | $p$ | $qr$ or $a$ | $\omega$ |
|---|---|---|---|---|---|---|---|---|---|---|
| VG | 0.319 | 0.060 | | 42.6 | 2.78 | | 1.04.E-3 | -0.01 | 2[b)] | |
| KBC | 0.320 | 0[b)] | 0.784 | 52.5 | 0.674 | 0.000 | 1.09.E-3 | 1.29 | 1.91 | |
| DBC | 0.323 | 0[b)] | 0.500 | 21.1 | 0.618 | 0.618 | 2.02.E-3 | 4.89 | 0.73 | |
| MDVC | 0.319 | 0[b)] | 0.813 | 42.6 | 2.78 | 1.00 | 1.13.E-3 | 1.13 | 1.67 | |
| PE | 0.320 | 0[b)] | 0.760 | 51.1 | 0.664 | | 1.44.E-3 | 0.30 | -2.64 | 8.60.E-4 |

Sand soil S 4660

| Model | $\theta_s$ | $\theta_r$ | $w$ | $H$ (cm) | $N_1$[a)] | $N_2$[a)] | $K_s$ (cm/s) | $p$ | $qr$ or $a$ | $\omega$ |
|---|---|---|---|---|---|---|---|---|---|---|
| VG | 0.436 | 0.049 | | 6.8 | 1.49 | | 1.45.E-2 | -0.70 | 2[b)] | |
| KBC | 0.450 | 0[b)] | 0.656 | 14.2 | 1.67 | 0.148 | 1.45.E-2 | 4.68 | 0.64 | |
| DBC | 0.438 | 0[b)] | 0.500 | 2.5 | 0.266 | 0.266 | 9.64.E-3 | 5.86 | 0.36 | |

| Model | $\theta_s$ | $\theta_r$ | $w$ | $H$ (cm) | $N_1$[a] | $N_2$[a] | $K_s$ (cm/s) | $p$ | $qr$ or $a$ | $\omega$ |
|---|---|---|---|---|---|---|---|---|---|---|
| MDVC | 0.436 | 0[b] | 0.888 | 6.8 | 1.49 | 1.00 | 6.88.E-3 | 5.20 | 0.61 | |
| PE | 0.451 | 0[b] | 0.749 | 17.1 | 1.68 | | 1.44.E-2 | 9.02 | -2.41 | 2.86.E-2 |

Sand soil S 4661

| Model | $\theta_s$ | $\theta_r$ | $w$ | $H$ (cm) | $N_1$[a] | $N_2$[a] | $K_s$ (cm/s) | $p$ | $qr$ or $a$ | $\omega$ |
|---|---|---|---|---|---|---|---|---|---|---|
| VG | 0.408 | 0.037 | | 9.9 | 1.82 | | 2.64.E-2 | -0.70 | 2[b] | |
| KBC | 0.416 | 0[b] | 0.798 | 16.6 | 1.21 | 0.132 | 2.64.E-2 | 3.14 | 0.97 | |
| DBC | 0.402 | 0[b] | 0.950 | 4.6 | 0.484 | 0.00 | 1.79.E-2 | 5.53 | 0.10 | |
| MDVC | 0.408 | 0[b] | 0.909 | 9.9 | 1.82 | 1.00 | 1.75.E-2 | 4.56 | 0.60 | |
| PE | 0.415 | 0[b] | 0.837 | 17.6 | 1.23 | | 2.39.E-2 | 0.97 | -1.68 | 2.70.E-4 |

Silty clay soil SiC 3120

| Model | $\theta_s$ | $\theta_r$ | $w$ | $H$ (cm) | $N_1$[a] | $N_2$[a] | $K_s$ (cm/s) | $p$ | $qr$ or $a$ | $\omega$ |
|---|---|---|---|---|---|---|---|---|---|---|
| VG | 0.641 | 0.000 | | 42.1 | 1.23 | | 9.75.E-5 | -0.70 | 2[b] | |
| MKBC | 0.673 | 0[b] | 1.000 | 857.2 | 3.30 | 205 | 8.83.E-5 | 6.00 | 0.50 | |
| DBC | 0.638 | 0[b] | 0.500 | 20.4 | 0.191 | 0.191 | 9.75.E-5 | 1.51 | 1.38 | |
| DVC | 0.641 | 0[b] | 0.787 | 42.1 | 1.23 | 1.23 | 9.72.E-5 | 6.76 | 0.38 | |
| MPE | 0.660 | 0[b] | 0.816 | 459 | 2.78 | | 5.67.E-5 | 9.95 | -2.25 | 5.44.E-4 |

Silty loam soil SiL 3370

| Model | $\theta_s$ | $\theta_r$ | $w$ | $H$ (cm) | $N_1$[a] | $N_2$[a] | $K_s$ (cm/s) | $p$ | $qr$ or $a$ | $\omega$ |
|---|---|---|---|---|---|---|---|---|---|---|
| VG | 0.310 | 0.000 | | 453 | 1.24 | | 1.37.E-5 | -0.70 | 2[b] | |
| MKBC | 0.328 | 0[b] | 0.961 | 6745.3 | 3.19 | 19200 | 1.41.E-5 | 0.71 | 1.34 | |
| DBC | 0.319 | 0[b] | 0.500 | 86.9 | 0.138 | 0.138 | 1.37.E-5 | 3.73 | 1.46 | |
| DVC | 0.310 | 0[b] | 0.780 | 453.0 | 1.24 | 1.24 | 1.37.E-5 | 0.30 | 1.82 | |

| | | | | | | | | | |
|---|---|---|---|---|---|---|---|---|---|
| MPE | 0.326 | 0[b] | 0.667 | 2000.0 | 2.53 | | 1.71.E-5 | 1.51 | -1.88 | 4.25.E-4 |

Silty loam soil SiL 3390

| Model | $\theta_s$ | $\theta_r$ | $w$ | $H$ (cm) | $N_1$[a] | $N_2$[a] | $K_s$ (cm/s) | $p$ | $qr$ or $a$ | $\omega$ |
|---|---|---|---|---|---|---|---|---|---|---|
| VG | 0.415 | 0.000 | | 33.9 | 1.19 | | 1.02.E-3 | -0.02 | 2[b] | |
| MKBC | 0.464 | 0[b] | 0.957 | 657.4 | 3.83 | 1.19 | 8.02.E-4 | 2.13 | 1.73 | |
| DBC | 0.400 | 0[b] | 0.500 | 20.5 | 0.153 | 0.153 | 6.52.E-5 | 0.56 | 2.03 | |
| DVC | 0.415 | 0[b] | 0.786 | 33.9 | 1.19 | 1.19 | 4.28.E-4 | 0.85 | 1.68 | |
| MPE | 0.445 | 0[b] | 0.783 | 340.3 | 3.00 | | 6.02.E-4 | 1.00 | -1.50 | 5.43.E-7 |

Silty loam soil SiL 3392

| Model | $\theta_s$ | $\theta_r$ | $w$ | $H$ (cm) | $N_1$[a] | $N_2$[a] | $K_s$ (cm/s) | $p$ | $qr$ or $a$ | $\omega$ |
|---|---|---|---|---|---|---|---|---|---|---|
| VG | 0.376 | 0.000 | | 148.2 | 1.16 | | 6.24.E-5 | 0.00 | 2[b] | |
| MKBC | 0.385 | 0[b] | 0.558 | 992.0 | 2.44 | 0.0449 | 9.52.E-6 | 0.31 | 1.40 | |
| DBC | 0.370 | 0[b] | 0.487 | 75.1 | 0.122 | 0.128 | 3.53.E-6 | 4.39 | 1.33 | |
| DVC | 0.376 | 0[b] | 0.800 | 148.2 | 1.16 | 1.16 | 4.50.E-6 | 4.07 | 1.00 | |
| MPE | 0.390 | 0[b] | 0.298 | 225.6 | 2.31 | | 3.85.E-5 | 1.00 | -1.50 | 4.68.E-4 |

Silty loam soil SiL 3393

| Model | $\theta_s$ | $\theta_r$ | $w$ | $H$ (cm) | $N_1$[a] | $N_2$[a] | $K_s$ (cm/s) | $p$ | $qr$ or $a$ | $\omega$ |
|---|---|---|---|---|---|---|---|---|---|---|
| VG | 0.355 | 0.000 | | 188.4 | 1.12 | | 6.42.E-5 | 0.00 | 2[b] | |
| MKBC | 0.371 | 0[b] | 0.771 | 9021.0 | 3.88 | 0.259 | 6.61.E-6 | 6.00 | 1.00 | |
| DBC | 0.355 | 0[b] | 0.502 | 71.5 | 0.091 | 0.091 | 4.22.E-6 | 0.70 | 1.78 | |
| DVC | 0.355 | 0[b] | 0.772 | 188.4 | 1.12 | 1.12 | 5.09.E-6 | 4.82 | 1.07 | |
| MPE | 0.358 | 0[b] | 0.617 | 4905.0 | 3.00 | | 1.96.E-5 | 0.72 | -3.41 | 2.35.E-5 |

Silty loam soil SiL 4673

| Model | $\theta_s$ | $\theta_r$ | $w$ | $H$ (cm) | $N_1$[a] | $N_2$[a] | $K_s$ (cm/s) | $p$ | $qr$ or $a$ | $\omega$ |
|---|---|---|---|---|---|---|---|---|---|---|
| VG | 0.401 | 0.113 |  | 575.1 | 1.75 |  | 4.76.E-5 | -0.04 | 2[b] |  |
| KBC | 0.407 | 0[b] | 0.354 | 547.3 | 1.56 | 0.224 | 5.65.E-5 | 0.45 | 1.72 |  |
| DBC | 0.400 | 0[b] | 0.500 | 198.0 | 0.248 | 0.248 | 4.76.E-5 | 3.88 | 1.76 |  |
| MDVC | 0.401 | 0[b] | 0.719 | 575.1 | 1.75 | 1.00 | 4.76.E-5 | 7.83 | 2.50 |  |
| PE | 0.403 | 0[b] | 0.546 | 858.1 | 1.16 |  | 4.76.E-5 | 7.44 | -1.34 | 9.17.E-4 |

Note: (a) $N_i = n, \lambda_i, n_i$ or $\sigma_i$ (b) Fixed parameter